\documentstyle[preprint,aps,eqsecnum,prd,epsf,amssymb]{revtex}

\def\spence{{\rm Li}}
\def\re{{\rm Re}\,}
\def\arcsinh{\,{\rm arcsinh}}
\def\arccosh{\,{\rm arccosh}}

\newcommand{\tr}[1]{{\rm tr}\left[#1\right]}
\def\MS{\overline{{\rm MS}}}

\def\dos{\delta_{\rm OS}}

\def\dlm{\delta_{\rm M}}

\begin{document}
\draft
\tighten

\title{
\vskip-3cm{\baselineskip14pt}
\centerline{\normalsize\hskip12.5cm hep-ph/9801265}
\centerline{\normalsize\hskip12.5cm DESY 97--259}
\centerline{\normalsize\hskip12.5cm HD--THEP 97--58}
\centerline{\normalsize\hskip12.5cm TUM--HEP--235}
\vskip1.5cm
The Higgs resonance in vector boson scattering
}
\author{
Wolfgang Kilian\thanks{\texttt{W.Kilian@thphys.uni-heidelberg.de}}$^1$
and
Kurt Riesselmann\thanks{\texttt{kurtr@ifh.de}}$^2$
}

\address{$^1$\,
 Institut f\"ur Theoretische Physik, Universit\"at Heidelberg,
 Philosophenweg 16\\
 D--69120 Heidelberg, Germany}

\address{$^2$\,
DESY Zeuthen, Platanenallee 6, D--15738 Zeuthen, Germany}

\maketitle
\vfill

\begin{abstract}
A heavy Higgs resonance is described in a representation-independent
way which is valid for the whole energy range of $2\to 2$ scattering
processes, including the asymptotic behavior at low and high energies.
The low-energy theorems which follow from to the custodial $SU_2$
symmetry of the Higgs sector restrict the possible parameterizations
of the lineshape that are consistent in perturbation theory.  Matching
conditions are specified which are necessary and sufficient to relate
the parameters arising in different expansions.  The construction is
performed explicitly up to next-to-leading order.
\end{abstract}

\pacs{PACS numbers: 14.80.Bn, 12.15.Lk, 11.10.St, 11.10.Hi}
\narrowtext

\section{Introduction}

Elastic scattering amplitudes of massive vector bosons grow
indefinitely with energy, if they are calculated perturbatively in a
theory of fermions and gauge bosons only.  As a result, the $S$-wave
scattering amplitudes of longitudinally polarized $W,Z$~bosons
manifestly violate unitarity beyond a critical energy scale
$\sqrt{s_c}\sim 1.2\;{\rm TeV}$~\cite{Uni}.

In the Minimal Standard Model (MSM)~\cite{SM}, an isodoublet of scalar
fields is introduced, leading to a single observable Higgs resonance
which damps the rise of those scattering
amplitudes~\cite{Higgs,SM-tree}.  However, the running Higgs
self-coupling~$\lambda$ increases with energy 
and becomes strong at some large scale~$\Lambda$ which
is indicated by a Landau pole in the one-loop running coupling
constant~\cite{LandauP}.  This scale depends exponentially on the
Higgs mass~$M_H$ and approaches the TeV range from above for
$M_H\gtrsim 400\ {\rm GeV}$.

Low-energy electroweak observables in the fermion/gauge boson sector
of the Standard Model are affected by radiative corrections which
depend logarithmically on~$M_H$.  From the high-precision data at
LEP1, SLC, and the Tevatron, an upper limit of $M_H<550\;{\rm GeV}$
has been derived at the $2\sigma$ level~\cite{MH-limits}.  This limit
is not sharp: Excluding one or two observables from the analysis
weakens the bound significantly~\cite{Baur}.  Furthermore, the limit
is strictly valid only within the context of the minimal model.  If
the Higgs mass is as large as several hundred GeV, effects from new
physics at the strong-interaction scale~$\Lambda$ could show up at low
energies in form of anomalous couplings.  Additional degrees of
freedom which invalidate the MSM calculation could also exist.

Thus, a Higgs resonance in the range $M_H=0.5\ldots 1\;{\rm TeV}$ is
still a realistic possibility.  Future colliders such as the LHC and
$e^+e^-$ linear colliders will explore the heavy-Higgs mass energy
range.  One expects that such a heavy Higgs resonance will be found if
it exists, and that from a precise analysis of its profile further
conclusions on the symmetry-breaking sector of the MSM can be drawn.
In order to separate anomalous effects, the predictions from the MSM
should be known as accurately as possible.  The results of this paper
are a step in this direction.

A heavy Higgs boson is not a quasi-stable particle that can safely be
treated in zero-width approximation.  Rather, the width of the Higgs
resonance will exceed~$100\ {\rm GeV}$ if $M_H\gtrsim 500\;{\rm GeV}$.
In a gauge theory the resummation of Feynman diagrams for an unstable
particle is not uniquely defined within the framework of perturbation
theory.  For the Higgs sector ambiguities arise when different
representations of the fields, which nevertheless are simply
related by field redefinitions, are compared.  However, amplitudes
for longitudinal $W,Z$~boson scattering have to satisfy low-energy
theorems~\cite{LET} analogous to those satisfied by pion scattering
amplitudes in low-energy QCD~\cite{sigma}, which in general are
violated if the Higgs width is introduced in a na\"\i{}ve way.

For many practical purposes, it may be sufficient to overcome these
problems by ad-hoc prescriptions.  Nevertheless, for a deeper
understanding, and if the theoretical predictions are to be used for
comparison with experiment, the uncertainties have to be under
control.  Therefore, we show in the present paper how different
descriptions and approximations valid in the low-energy and
high-energy ranges and in the resonance region can be combined to
yield a unified resummation prescription which is valid within the
whole perturbative range.  Applying matching and resummation
procedures consistently, representation and renormalization-scheme
dependence disappears order by order in the perturbative expansion.

The paper is organized as follows: In the following two sections we
introduce the Higgs resonance in the context of Goldstone boson
scattering and discuss different representations of the Higgs and
Goldstone fields in the Lagrangian.  In particular, we give a
one-parameter formula which interpolates between the linear and
non-linear representations.  In Sec.\ref{sec:reson} we state
the conditions that have to be imposed on the Goldstone boson
scattering amplitude, and in Sec.\ref{sec:resum} we show how they
resolve the apparent ambiguities which arise when the finite width of
the Higgs boson is taken into account.  The next three sections are
devoted to the explicit calculation of the Higgs lineshape at leading
and next-to-leading order.  Finally, we discuss representation
dependence and its use for estimating higher-order effects and conclude.

\section{Theoretical framework}
\label{sec:theory}
For a quantitative analysis, the Higgs lineshape should be calculated
within the full MSM.  The physical picture, however, is clearer if
only the leading contributions for high energies and for a large Higgs
mass are taken into account.  For this reason, we shall discuss the
Higgs resonance within the framework of the Equivalence Theorem
(ET)~\cite{SM-tree,ET,He,ETadd} which relates the unphysical Goldstone
modes in a $R_\xi$ gauge to the longitudinal degrees of freedom of the
$W,Z$ bosons in unitary gauge.  In this approximation, one
consistently neglects terms of order $g^2$ compared to those of order
$g^2M_H^2/M_W^2$ or $g^2 s/M_W^2$.  

Corrections induced by the top quark Yukawa coupling are smaller than
Higgs coupling corrections if $M_H\gtrsim 2m_t$~\cite{DV}.  We thus
neglect them in the present paper, deferring their inclusion to a
future publication.

If the Yukawa couplings are set to zero, the theory has a global
$SU_2\times SU_2$ symmetry~\cite{SU2c} that is spontaneously broken to
the diagonal $SU_2$.  From this fact one can conclude~\cite{sigma}:
(i) The three Goldstone modes can be kept massless consistently to all
orders. (ii) For all $2\to 2$ Goldstone scattering amplitudes, the
dynamical dependence on the Mandelstam variables $s,t,u$ is determined
by a single function
\begin{equation}
  A(s,t,u) = A(s,u,t)
\end{equation}
which is equal to the amplitude for $ww\to zz$ scattering.  Here
$\{w,z\}$ denote the Goldstone modes associated with the
longitudinally polarized $W,Z$ bosons. (iii) As $s$~goes to zero, the
real part of the amplitude $A$ vanishes like~$s$, and the imaginary
part which arises at one-loop order vanishes like~$s^2$.  In this
limit both terms are determined completely by low-energy quantities,
independent of the existence of a Higgs resonance.  (iv) For very high
energies, the Higgs mass can be neglected, and the theory approaches a
massless $\phi^4$ theory.  In this limit the $SU_2\times SU_2$
symmetry is manifest in the scattering diagrams; it is only slightly
broken by the renormalization conditions necessary to match the
asymptotic behavior with the low-energy theory, as will be described
below.

\section{Representations of the Higgs sector}
\label{sec:rep}
The Lagrangian describing the MSM Higgs sector in the absence of
Yukawa and gauge couplings reads
\begin{equation}\label{sigma-L}
  {\cal L} = \frac12\tr{\partial_\mu\Sigma^\dagger\partial^\mu\Sigma}
             + \frac{M^2}{4}\tr{\Sigma^\dagger\Sigma}
	     - \frac{M^2}{8v^2}\left(\tr{\Sigma^\dagger\Sigma}\right)^2,
\end{equation}
Here $\Sigma$ is a $2\times 2$ matrix that transforms under
$SU_2\times SU_2$ as $\Sigma \to V^\dagger \Sigma U$ and has a vacuum
expectation value $\langle\Sigma\rangle = v/\sqrt{2}$.  It may be
parameterized in terms of four real scalar fields $H$ and $w_a$
($i=1,2,3$) as
\begin{equation}\label{linear}
  \Sigma = \frac{1}{\sqrt2}\left[\left(v + H\right){\bf 1}
	+ iw_a\tau^a\right]
\end{equation}
where $\tau^a$ are the Pauli matrices.  In this parameterization the
symmetry is represented linearly, and renormalizability ---
\emph{i.e.}, the logarithmic high-energy behavior --- is manifest.

On the other hand, a non--linear representation
\begin{equation}\label{nonlinear}
  \Sigma = \frac{1}{\sqrt2}(v+H){\bf 1}
	\exp\left(\frac{i}{v}w'_a\tau^a\right)
\end{equation}
may be preferred for the description of low--energy scattering
amplitudes, since the Higgs field can be integrated out to leading
order by letting $H\to 0$, resulting in a non-linear $\sigma$~model
where the power--like low--energy behavior is manifest.  Furthermore,
in this representation individual terms can be identified more
easily in the full unitary-gauge MSM amplitudes, since the derivative
couplings in the Goldstone interactions correspond to the momentum
factors in the longitudinal polarization vectors of massive vector
bosons.

The physics derived from a Lagrangian, however, is independent of the
particular representation chosen for the fields; it depends only on
the number of degrees of freedom, their symmetry properties, and on
numerical parameters.  In fact, $S$--matrix amplitudes --- as opposed
to off-shell Greens functions --- are invariant under a wide class of
non-linear field transformations, if the independent parameters of the
theory are expressed in terms of measurable
quantities~\cite{HRB,CCWZ,BL,BM}.  The expansion of the fields
$\{w'_a,H\}$ in terms of $\{w_a,H\}$, and vice versa, can easily be
worked out order by order.  Hence, the linear and the non-linear
representations yield the same results if \emph{all} Feynman diagrams
to a given order in the perturbative expansion are taken into account.

In order to make representation dependence explicit, one may introduce
a one-parameter family of field representations which includes both
representations mentioned above:
\begin{equation}\label{eta-rep}
  \Sigma = \frac{1}{\sqrt2}\left[\left(\frac{v}{\eta} + H\right)
		\exp\left(i\frac{\eta}{v}w''_a\tau^a\right)
		+ \left(1-\frac{1}{\eta}\right)v\right].
\end{equation}
Here $\eta$ is an arbitrary parameter.  Taking $\eta\to 0$, the linear
representation~(\ref{linear}) is reproduced. The choice $\eta=1$
corresponds to the non-linear representation~(\ref{nonlinear}).  If
the Feynman rules derived by inserting~(\ref{eta-rep}) into the
Lagrangian~(\ref{sigma-L}) are used to calculate a physical quantity,
the parameter~$\eta$ must drop out of the result if all diagrams up to
a given order of the expansion parameter chosen are taken into
account.  In this sense the parameter~$\eta$ achieves a similar
meaning as the gauge parameter~$\xi$ of the electroweak theory.  We
will come back to this fact in Sec.\ref{sec:eta}.

\section{The Higgs resonance in $ww\to zz$}
\label{sec:reson}
\begin{figure}
\begin{center}\leavevmode
\epsffile{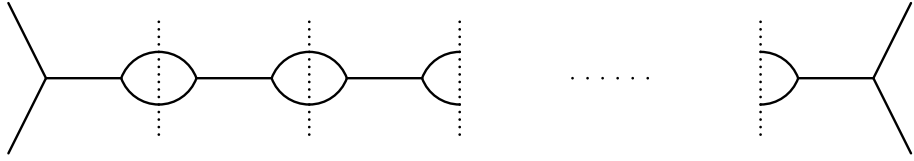}
\end{center}
\caption{Dyson resummation of resonant terms in Goldstone scattering.
The cuts indicate where the imaginary part of the loop is taken.}
\label{fig:Dyson}
\end{figure}

At lowest order, the amplitude for $ww\to zz$ scattering is easily
calculated as
\begin{equation}\label{A-LO}
  A^{(0)}(s,t,u) \equiv A^{(0)}(s) = -2\lambda\frac{s}{s-M^2}
\end{equation}
with $\lambda=M^2/2v^2$.  As anticipated, this expression does not
depend on~$\eta$.  It satisfies the requirements mentioned in
Sec.\ref{sec:theory}:
\begin{itemize}
\item[(a)] At low energies, it approaches the expression $s/v^2$.
Generally speaking, the sum of the diagrams in any fixed order $n$ of
the perturbative expansion vanishes like $s^{n+1}/(4\pi v)^{2n}\log^n
s$ for $s\to 0$.  Thus, the leading-order low-energy behavior
is not modified by higher-order corrections.
\item[(b)] At high energies, the amplitude $A^{(0)}(s)$ approaches a
constant value.  Higher-order corrections modify this by adding
logarithmic terms of order $\lambda^n\ln^k s$, $k\leq n$.
\end{itemize}
However, this amplitude cannot be used in the resonance region since
it diverges at $s=M^2$.  Resonant diagrams need to be resummed,
leading to the introduction of the Higgs width which to leading order
is given by
\begin{equation}
  \Gamma^{(0)} = \frac{3\lambda}{16\pi}M.
\end{equation}
Several approaches to deal with this problem are
possible~\cite{Zres,LEP1,Zres-S,Hres,BHKO,pinch}.  Let us list some
of them as they are applied to the leading-order
expression~(\ref{A-LO}):
\begin{enumerate}
\item \textbf{$S$-matrix approach.}  The Higgs pole term is separated
with a constant residue, and the correct pole position in the complex
plane is inserted.  The remainder, the non-resonant part, is left
untouched at this order:
\begin{equation}
  A_1^{(0)}(s,t,u)\equiv A_1^{(0)}(s) 
	= \frac{M^2}{s-M^2+iM\Gamma\theta(s)} + 1
\end{equation}
The constant width $\Gamma$ is only introduced for $s>0$, as indicated
by the $\theta$~function.
\item \textbf{Fixed width in common denominator.}  All terms,
including the non-resonant part, are collected into a single fraction
before the pole position in the denominator is shifted:
\begin{equation}
  A_2^{(0)}(s) = -2\lambda\frac{s}{s - M^2 + i M\Gamma\,\theta(s)}
\end{equation}
\item \textbf{Resummation of self-energies.}  The separation is done
by grouping Feynman diagrams in two classes: resonant and
non-resonant.  In the resonant diagrams~(Fig.\ref{fig:Dyson}) the
imaginary parts of the one-particle irreducible piece (the Higgs
self-energy) are resummed\footnote{At this order, this is equivalent
to resumming the full self-energy.}.  If the $s$-dependence is kept,
one finds
\begin{equation}\label{decomp-eta}
  A_3^{(0)}(s) = -2\lambda\frac{1}{M^2}\,
	\frac{[\eta s + (1-\eta)M^2]^2}{s - M^2 + iM\Gamma(s)\,\theta(s)}
	+ 2\lambda\frac{\eta^2 s - (1-\eta)^2M^2}{M^2}
\end{equation}
where the $s$-dependent width is also representation dependent,
\begin{equation}\label{Gamma(eta)}
  \Gamma(s) = \frac{3\lambda}{16\pi M^3}[\eta s + (1-\eta)M^2]^2.
\end{equation}
\item \textbf{Kinematical scaling of a constant width $\Gamma$.}  The
phase space available for the decay of a state with mass $M^*=\sqrt{s}$
into massless particles scales proportional to~$s$.  Applying this
observation to the intermediate Higgs state in Goldstone scattering,
one is led to
\begin{equation}
  A_4^{(0)}(s) = \frac{s}{s-M^2+is\Gamma\theta(s)/M}.
\end{equation}
where $\Gamma$ itself is independent of $s$.
\end{enumerate}
\emph{A priori}, no single approach is preferred.  In principle, the
inclusion of higher-order corrections can be done as to remove the
discrepancies between different formulae.  However, the differences
can be numerically large at tree-level.  This has been explicitly
verified for the analogous problem of the $Z$ and
$W$~resonances~\cite{Zres-S,Wres}.  [Note that for the $Z$~resonance,
the kinematical-scaling and self-energy resummation schemes,
approaches (3) and (4), coincide in a linear gauge.]

Considering the Higgs resonance, we observe that the expressions
$A_1^{(0)}$ to~$A_3^{(0)}$ are not in accordance with the low-energy
theorem.  The amplitude~$A_1^{(0)}$ approaches a constant for $s\to
0$.  The version~$A_2^{(0)}$ has the correct power behavior, but the
normalization is changed to an unphysical complex value.  [Recall that
for $s\to 0$ an imaginary part is allowed only at order~$s^2$.]  Even
worse, the expression $A_3^{(0)}$ depends on~$\eta$, \emph{i.e.}, on
the representation chosen for the Higgs sector.  By contrast, the
expression~$A_4^{(0)}$ which is seemingly introduced by an \emph{ad
hoc} replacement, satisfies both low-energy and
representation-independence requirements.

Apparently, the ambiguity in finding a correct resonance description
can be removed by the condition that the low-energy theorem should be
maintained to all orders\footnote{For the lowest-order expression,
this has been shown in~\cite{Sey}}.  This additional requirement,
which has no counterpart for the $Z$~resonance, restricts the possible
parameterizations of the Higgs resonance.  As will be demonstrated in
the following sections, imposing the low-energy theorem on the
amplitude in each order of the perturbative expansion unites the
$S$-matrix approach with the concept of kinematical scaling.
Furthermore, it provides matching conditions which fix the
renormalization of the independent parameters.  On the other hand, a
scheme based on the representation-dependent expression~$A_3$ can be
set up to give identical results, as shown in Sec.\ref{sec:eta}.

\section{Representation independent treatment of the Higgs resonance}
\label{sec:resum}
In perturbation theory the property of a Feynman diagram to be
one-particle reducible or irreducible is not well defined; it depends
on the particular representation chosen for the fields~[cf.\
(\ref{decomp-eta})].  Therefore, a resummation of ``resonant''
diagrams is ambiguous in general.  This observation naturally leads to
a parameterization of a resonance inspired by $S$-matrix
theory~\cite{Zres-S}, which we will adopt as a starting point for the
treatment of the Higgs lineshape.  However, as we have seen in the
previous section, for the Higgs resonance this approach is not
manifestly consistent with the low-energy theorem, if finite-order
perturbation theory is applied.  For this reason, we first review the
various perturbative expansions and the conditions they have to
satisfy.  In particular, four energy ranges with different expansion
parameters have to be distinguished:
\begin{enumerate}
\item The low-energy region ($s\ll M^2$): The amplitude is expanded
in powers of $s/(4\pi v)^2$.  The leading term and the imaginary part
of the next-to-leading term are fixed by the low-energy theorem.
\item The perturbative region ($s\sim M^2$, but excluding the
resonance region where $|s-M^2| \lesssim \lambda M^2/16\pi$): The amplitude is
expanded in powers of $\lambda/16\pi = M^2/32\pi v^2$.
\item The resonance region ($|s-M^2| \lesssim\lambda M^2/16\pi$): The
distance from the pole is of the order of the width
$\Gamma\sim\lambda$.  Any Feynman diagram contributing to $ww\to zz$
can be characterized by non-negative integers $n$ and $k$ to be of the
order
\begin{equation}\label{res-representation}
  16\pi\left(\frac{\lambda}{16\pi}\right)^n 
  \left(\frac{\lambda M^2/16\pi}{s-M^2}\right)^k,
  \qquad n\geq 1, \quad k\geq 0,
\end{equation}
where $k$ counts the number of resonant propagators, and all
$s$-dependence that is not determined by the pole terms has been
absorbed in the $k=0$ piece.  All terms with a fixed $n$ are formally
of the same order and need to be resummed.
\item The high-energy region ($s\gg M^2$): Neglecting everything that
is suppressed by powers of $M^2/s$, any term can be assigned
non-negative integers $n$ and $k$ to be of the order
\begin{equation}
  16\pi\left(\frac{\lambda}{16\pi}\right)^n
  \left(\frac{\lambda}{16\pi}\ln\frac{s}{M^2}\right)^k,
  \qquad n\geq 1, \quad k\geq 0.
\end{equation}
All terms with a fixed $n$ are formally of the same order and need to
be resummed.  This can be accomplished by renormalization-group
methods, introducing a running coupling and field anomalous
dimensions.
\end{enumerate}
In each of these expansions the individual terms are
representation-independent: For the perturbative expansion
in~$\lambda$, this follows from the general theorems mentioned above.
Regarding the expansion in powers and logarithms of~$s$, the
corresponding pieces can in principle be identified in a measurement
of physical scattering processes.

Let us first look at the resonance region.  The full $S$-matrix has no
multiple poles, but only a simple pole at the complex position
$s_P=M^2-iM\Gamma$.  This defines two real representation-independent
constants $M$ and $\Gamma$, which can be interpreted as the physical
mass and physical width of the Higgs resonance\footnote{The pole
position can be determined from the Higgs self-energy by solving an
implicit equation.  One has to be careful, however, since the
self-energy by itself is representation-dependent, and the extraction
of $M$ and $\Gamma$ from the corresponding Feynman amplitudes can be
technically quite involved at higher orders.}~\cite{Pole}.  From
re-expanding the complex pole around $s=M^2$ we recover the
expansion~(\ref{res-representation}), and we observe that resonant
terms, \emph{i.e.} those with order $k\geq 1$, can be summed up into
\begin{eqnarray}\label{Dyson-res}
  A_{\rm res}(s,t,u) &=& 
	(\alpha+i\beta)\frac{M^2}{s-M^2}\sum_{k=1}^\infty
	\left(-i\frac{\Gamma}{M}\theta(s)\frac{M^2}{s-M^2}\right)^{k-1}
	\nonumber\\
       &=& (\alpha+i\beta)\frac{M^2}{s-M^2(1-i\gamma)},
\end{eqnarray}
where $\gamma\equiv (\Gamma/M)\theta(s)$.  This is the usual Dyson
series.  Assuming that any field-strength renormalization of the
external Goldstone particles is included in the amplitude $A(s,t,u)$,
the relation~(\ref{Dyson-res}) defines two more real observable
quantities $\alpha$ and $\beta$.  If the definition of the expansion
parameter~$\lambda$ is given in terms of physical observables, the
constants $\alpha$, $\beta$, and $\gamma$ have perturbative expansions
that are independent of the representation and of any intermediate
renormalization scheme order by order.

The remaining part of the full amplitude $A(s,t,u)$ can be collected
in a function $A_{\rm nr}$
which we may call the non-resonant piece.  The result for the total
amplitude is
\begin{eqnarray}\label{Dyson}
  A(s,t,u) &=& (\alpha+i\beta)\frac{M^2}{s-M^2(1-i\gamma)} 
	+ A_{\rm nr}(s,t,u).
\end{eqnarray}
With these definitions the function~$A_{\rm nr}$ also has a
perturbation expansion in which each term is representation- and
scheme-independent separately, and its higher-order terms scale like
$s^{n+1}\ln^n s$ in the low-energy limit.

In a real calculation the quantities $\alpha$, $\beta$, $\gamma$, and
$A_{\rm nr}$ can be computed only to finite order.  It is this
truncation which spoils the low-energy theorem, as we have observed in
the lowest-order example of the preceding section.  However, we are
free to rewrite the exact expression~(\ref{Dyson}) in such a way that
the correct low-energy behavior is preserved even for the truncated
series.  To this end, we evaluate~(\ref{Dyson}) for $s=0$:
\begin{equation}
  A(0) = 0 = -\frac{\alpha+i\beta}{1-i\gamma} 
	+ A_{\rm nr}(0)
\end{equation}
Thus, the modified function
\begin{equation}\label{A-hat}
  \hat A_{\rm nr}(s,t,u) \equiv 
  A_{\rm nr}(s,t,u) - \frac{\alpha+i\beta}{1-i\gamma} 
\end{equation}
vanishes like $s$ for $s\to 0$.  

It remains to introduce the piece~(\ref{A-hat}) in the complete
expression~(\ref{Dyson}).  Rewriting~(\ref{Dyson}), we obtain
\begin{eqnarray}\label{ImpDyson}
  A(s,t,u) &=& (\alpha+i\beta)\frac{1+i\gamma}{1-i\gamma}\,
	\frac{s}{s-M^2(1+\gamma^2) + i\gamma s} 
	+ \hat A_{\rm nr}(s,t,u)
\end{eqnarray}
to all orders.  [Recall that $\gamma=(\Gamma/M)\theta(s)$ with a
constant width~$\Gamma$.]

By extracting the factor~$1+i\gamma$ in the numerator, we have
effectively introduced a kinematical scaling proportionaly to~$s$ of
the constant Higgs width in the denominator.  When the result is
re-expanded in the low-energy region, the imaginary part of the
denominator enters only at order~$s^2$, allowing the imaginary parts
of the first and second term to cancel up to order~$s$ at each order
in the loop expansion.  This is necessary to satisfy the low-energy
theorem order by order.  As we shall see later, the \emph{real} part
of the order-$s$ term is divergent; here the low-energy theorem serves
as a matching condition
\begin{equation}\label{LE-matching}
  A(s,t,u) \stackrel{s\to 0}{\rightarrow} \frac{s}{v^2} + {\cal O}(s^2),
\end{equation}
which can be employed to define the renormalized coupling~$\lambda$ in
terms of $M$ and the low-energy quantity $v\equiv(\sqrt{2}\,G_F)^{-1/2}$.

It is easily verified that the result~(\ref{ImpDyson}) is correct in
the other energy ranges: At $s=M^2$, terms have been shifted between
the resonant and the non-resonant piece.  However, if the width is
calculated to order $n$, one has to calculate $\hat A_{\rm nr}$ only
up to order $n-1$ to ensure that the normalization of the peak
amplitude is unchanged.  For $s>M^2$ the resonant part
in~(\ref{Dyson}) vanishes like $1/s$, and only $A_{\rm nr}$ is of
importance.  In~(\ref{ImpDyson}) both parts contribute for
$s\to\infty$.  However, since there is no cancellation required,
this fact is irrelevant.

We have not yet considered the resummation of logarithms in the
high-energy limit.  For $s\gg M^2$, the theory eventually approaches a
massless $\phi^4$ model.  This fact can be employed to include those
terms of order $\lambda^n\ln^k s$ in the amplitude that are
not already part of the finite-order expression.  With our
prescription for the treatment of the resonance, such logarithmic
terms cannot be picked up by the pole resummation implicit
in~(\ref{ImpDyson}).  Thus, they can be added separately while
avoiding any double-counting, leading to the final result
\begin{eqnarray}\label{amp-full}
  A_{\rm full}(s,t,u) &=& 
	(\alpha+i\beta)\frac{1+i\gamma}{1-i\gamma}\,
	\frac{s}{s-M^2(1+\gamma^2) + i\gamma s} 
	+ \hat A_{\rm nr}(s,t,u)
	\nonumber\\
  && {} + \left[A_{\rm RG}(s,t,u;\mu_0^2) 
	- A_{\rm HE}(s,t,u;\mu_0^2)\right]\theta(s,t,u;\mu_0^2),
\end{eqnarray}
where the first two terms are taken from~(\ref{ImpDyson}).  If the
amplitude $A(s,t,u)$ were known to all orders, the additional term
introduced here would vanish identically, and there would be no
dependence on the matching point~$\mu_0$.  However, in a finite-order
calculation the correction $A_{\rm RG}-A_{\rm HE}$ provides just those
logarithmic terms that are not included already in~(\ref{ImpDyson})
and which dominate in the high-energy limit.  To achieve this, we
define the functions $\theta$, $A_{\rm RG}$, and $A_{\rm HE}$ as
follows:

The step function $\theta(s,t,u;\mu_0^2)$ controls the region where
resummation of logarithms applies.  A natural choice is\footnote{In
the forward and backward regions where $|t|\ll s$ or $|u|\ll s$, a
straightforward resummation of logarithms~$\ln s$ does not pick up the
dominant terms.  To exclude these regions from the
renormalization-group improvement, one should insert additional
cutoff functions in~$t$ and $u$:
\begin{displaymath}
  \theta(s,t,u;\mu_0^2) = \theta(s-\mu_0^2)\,
	\theta(-t-\mu_0^2/2)\,\theta(-u-\mu_0^2/2).
\end{displaymath}
This definition has been used in our numerical results for the
eigenamplitude~$a_{00}$ (cf.\ Sec.\ref{sec:discussion}).  For the
presentation of differential distributions, some smooth cutoff in~$t$
and $u$ would be more appropriate.  However, the effect of this
ambiguity on the total cross section is suppressed by~$1/s$.}
\begin{equation}\label{theta}
  \theta(s,t,u;\mu_0^2) = \theta(s-\mu_0^2).
\end{equation}
with a matching point $\mu_0\sim M$.

To define the amplitude $A_{\rm HE}$ in~(\ref{amp-full}), we
observe that the finite-order expression $A(s,t,u)$ in~(\ref{Dyson}),
or, equivalently, the resummed expression~(\ref{ImpDyson}), has an
asymptotic form:
\begin{equation}
  A_{\rm he}(s,t,u;M^2) = \lim_{s\gg M^2}A(s,t,u).
\end{equation}
The limit is to be taken such that constant terms and logarithms
$\ln^k s$, $\ln^k u$, etc.\ are kept, but all terms suppressed by at
least one power of $1/s$ are omitted.  The result depends on the
Mandelstam variables $s,t,u$ and on a mass scale which can be chosen
as the pole mass~$M$.  

On the other hand, the amplitude can be calculated in the high-energy
effective theory, \emph{i.e.}, in the massless $\phi^4$~model.  The
result, $A_{\rm HE}(s,t,u;\mu^2)$, depends on an arbitrary mass
scale~$\mu$.  If it is to represent the high-energy limit of the
amplitude~$A(s,t,u)$, it must satisfy
\begin{equation}\label{HE-matching}
  A_{\rm he}(s,t,u;M^2) = A_{\rm HE}(s,t,u;\mu_0^2),
\end{equation}
where~$\mu$ is identified with the matching scale~$\mu_0$.  If some
intermediate renormalization scheme (\emph{e.g.}, $\MS$) is used to
define the high-energy effective theory, matching corrections must be
added order by order such that~(\ref{HE-matching}) is fulfilled.

Finally, the amplitude $A_{\rm RG}(s,t,u)$ is derived from the
expression $A_{\rm HE}(s,t,u)$ by standard renormalization-group
methods: A running coupling~$\lambda(s)$ and field anomalous
dimensions are introduced to absorb the logarithms $\ln^k(s/\mu_0^2)$
order by order.  The initial condition is
\begin{equation}\label{RG-matching}
  A_{\rm RG}(s,t,u;\mu_0^2) =
  \left. A_{\rm HE}(s,t,u;\mu_0^2) \right|_{s=\mu_0^2}
\end{equation}
at the matching point $s=\mu_0^2$.  This fixes all parameters of the
high-energy effective theory and their renormalization in terms of $M$
and~$\lambda$.  Including the matching of the low-energy effective
theory~(\ref{LE-matching}), the only two independent parameters are
$M$ and~$G_F$.

Our prescription for the treatment of the Higgs resonance is not
unique.  If all requirements are maintained that have to be imposed on
the scattering amplitude, additional terms can be shifted from the
non-resonant into the resonant part~\cite{pinch}.  The difference is
then subleading in the respective expansion parameter for all energy
ranges, and eventually disappears if all orders are taken into
account.  However, if logarithmic terms are shifted into the resonant
piece which is resummed in~(\ref{Dyson-res}), they become relevant in
the high-energy limit, and one has to be careful to correctly include
them in the matching conditions for the renormalization-group improved
result.

\section{Lowest-order result}
\label{sec:LO}
Considering the expansion of~(\ref{ImpDyson}) in powers of $\lambda$,
we note that
\begin{equation}\label{reexpand}
  (\alpha+i\beta)\frac{1+i\gamma}{1-i\gamma}
	= \alpha^{(0)} + \alpha^{(1)} 
		+ i\left(\beta^{(1)}+2\alpha^{(0)}\gamma^{(0)}\right) + \ldots,
\end{equation}
where $\alpha^{(0)}$, $\gamma^{(0)}\propto\lambda$.  Keeping only the
leading term:
\begin{equation}
  \alpha^{(0)}=A_{\rm nr}^{(0)}=-2\lambda.
\end{equation}
To leading order, the general expression~(\ref{ImpDyson}) reduces
to~\cite{Sey}
\begin{eqnarray}\label{Seymour}
  A^{(0)}(s) 
  &=& -2\lambda\frac{s}{s-M^2 + i\frac{s}{M}\Gamma^{(0)}\theta(s)}.
\end{eqnarray}
Here the non-resonant piece $\hat A_{\rm nr}$ vanishes identically,
Re-expanding for $s\ll M^2$, the imaginary part of order~$s^2$ is
reproduced correctly, satisfying the low-energy theorem.

\begin{figure}
\begin{center}\leavevmode
\epsffile{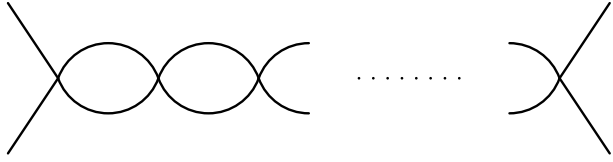}
\end{center}
\caption{The set of one-loop diagrams relevant in the high-energy limit}
\label{fig:LLA}
\end{figure}

In the high-energy region renormalization group improvement is
necessary.  The leading logarithmic approximation is obtained by
resumming the energy dependence of the bubbles in Fig.\ref{fig:LLA}
in the massless limit, using a running coupling,
\begin{equation}\label{LLA}
  A_{\rm RG}^{(0)}(s) = -2\lambda^{(0)}(s),
\end{equation}
where the running coupling is
\begin{equation}
  \lambda^{(0)}(\mu^2) = 
  \frac{\lambda}{1 - \beta_0\lambda\ln\frac{\mu^2}{\mu_0^2}}
\end{equation}
with a matching point $\mu_0\sim M$, and
\begin{equation}
  \beta_0 = 12/16\pi^2.
\end{equation}

If we want to implement~(\ref{LLA}) in our previous result for
$A^{(0)}(s)$, we have to subtract the high-energy limit
of~(\ref{Seymour})
\begin{equation}
  A^{(0)}(s) \stackrel{s\gg M^2}{\longrightarrow} A^{(0)}_{\rm he}(s) 
	= A^{(0)}_{\rm HE}(s) = - 2\lambda.
\end{equation}
Thus we obtain the final result~(\ref{amp-full}) at leading order
\begin{eqnarray}\label{A0-full}
  A^{(0)}_{\rm full}(s) 
	&=& A^{(0)}(s) 
	+ \left[A^{(0)}_{\rm RG}(s)-A^{(0)}_{\rm HE}(s)\right]
	\theta(s,t,u;\mu_0^2)
	\nonumber\\
	&=& -2\lambda\frac{s}{s-M^2 + i\frac{s}{M}\Gamma\theta(s)}
	-2\lambda
	\frac{\beta_0\lambda\ln\frac{s}{\mu_0^2}}
	{1 -\beta_0\lambda\ln\frac{s}{\mu_0^2}}\theta(s,t,u;\mu_0^2).
\end{eqnarray}
The step function $\theta(s,t,u;\mu_0^2)$ which controls the region of
renormalization-group improvement is defined in~(\ref{theta}).

\section{One-loop result}
\label{sec:NLO}
If loop diagrams are taken into account, one usually has to specify a
renormalization scheme.  In our approach this is unnecessary in
principle: We could work with the dimensionally regularized, but
unrenormalized expressions.  Matching the amplitudes between the
different energy regions is equivalent to a complete set of
renormalization conditions.  However, any intermediate renormalization
scheme eventually leads to the same physical amplitudes to the order
considered: For illustration, we present our results both in the
on-shell ($\dos=1$) and in the $\MS$ ($\dos=0$) schemes.  In the
latter scheme, the scale~$\mu$ is set equal to the $\MS$ Higgs
mass~$m$, unless stated differently.  All relevant diagrams have been
calculated in Refs.\cite{ChPT,GH-NLO,NLO,GJR}; we will use the
notation of Ref.\cite{NLO} with slight modifications.  The functions
and constants introduced below are defined in the Appendix.

As before, we define $M^2$ to be the real part of the pole position.
For the one-loop matching we need its value in terms of the $\MS$
renormalized mass $m$ to one-loop order:
\begin{equation}\label{m-matching}
  M^2 = m^2\left(1 + \delta m^2/m^2\right)
\end{equation}
with
\begin{equation}
   \frac{\delta m^2}{m^2} = -\frac{\lambda}{16\pi^2}
  \left(24 - 9K_1\right)(1-\dos).
\end{equation}
In the on-shell scheme, $m^2=M^2$ by definition.

\begin{figure}
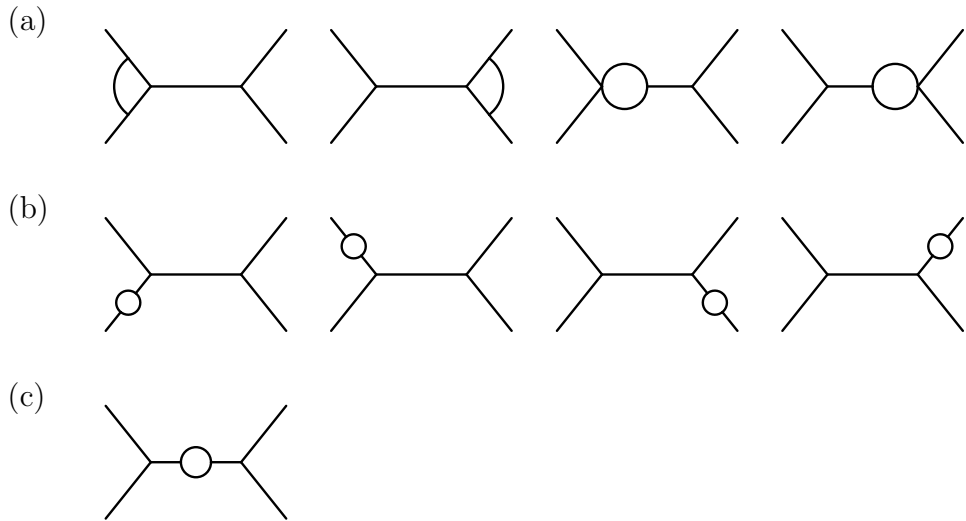

\begin{center}\leavevmode
\unitlength 1mm
\begin{picture}(110,88)
\put(0,0){\epsffile{fmgraph.14}}
\put(0,25){\epsffile{fmgraph.10}}
\put(30,25){\epsffile{fmgraph.11}}
\put(60,25){\epsffile{fmgraph.12}}
\put(90,25){\epsffile{fmgraph.13}}
\put(0,50){\epsffile{fmgraph.2}}
\put(30,50){\epsffile{fmgraph.3}}
\put(60,50){\epsffile{fmgraph.8}}
\put(90,50){\epsffile{fmgraph.9}}
\put(-10,65){(a)}
\put(-10,40){(b)}
\put(-10,15){(c)}
\end{picture}
\end{center}
\caption{Resonant one-loop Feynman diagrams for Goldstone scattering.
By repeating these diagrams as in Fig.\ref{fig:Dyson} the one-loop
corrected width is generated in the Higgs propagator.}
\label{fig:1L-resonant}
\end{figure}

If expressed in terms of $M^2$, the leading-order (LO) resonant
diagrams have the structure depicted in Fig.\ref{fig:Dyson}, where a
singly resonant amplitude is repeated an infinite number of times
between two-particle cuts.  In the first set of next-to-leading order
(NLO) diagrams, one of those singly resonant parts has one additional
power of $\lambda$.  We therefore need the residue of the pole at
$s=M^2$ of the diagrams in Fig.\ref{fig:1L-resonant}.  The remainder of
the expansion around $s=M^2$ is considered non-resonant and will be
taken into account later.

For the diagrams in Fig.\ref{fig:1L-resonant}b we observe that the
self-energy insertion on the external lines can occur both to the
right and to the left of a two-particle cut, so that it should be
counted with a factor $1/2$.  In the diagram~\ref{fig:1L-resonant}c,
which is double resonant, the singly resonant part is obtained from
the $s-M^2$ term in the Taylor expansion of the Higgs self-energy.
The real part of the sum of the diagrams in Fig.\ref{fig:1L-resonant}
determines the imaginary part of the pole position,
$M\Gamma^{(1)}$~\cite{GH-NLO}, expressed in a particular
renormalization scheme:
\begin{equation}
  M\Gamma^{(1)} = M^2\frac{3\lambda}{16\pi}
	\left(1 + \gamma_1\right)
\end{equation}
with
\begin{equation}
  \gamma_1 = \frac{\lambda}{16\pi^2}\left[
	-15-3K_1+4G_1 + 12H_1 - 9K_1'
	+ \delta_{\rm OS}\left(25 - 9K_1\right)\right].
\end{equation}

\begin{figure}
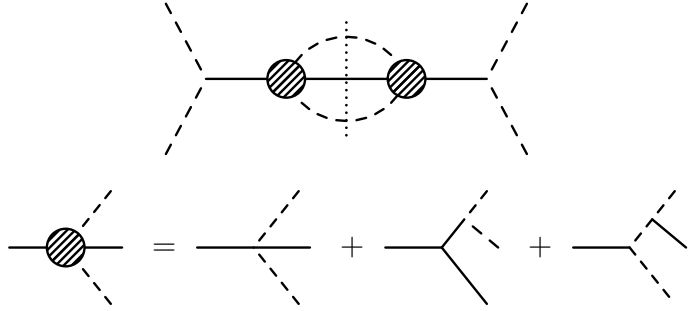

\begin{center}\leavevmode\epsffile{fmgraph.30}\end{center}
\begin{center}\leavevmode
\unitlength 1mm
\begin{picture}(90,15)
  \put(0,0){\epsffile{fmgraph.31}}
  \put(19,6.5){$=$}
  \put(25,0){\epsffile{fmgraph.32}}
  \put(44,6.5){$+$}
  \put(50,0){\epsffile{fmgraph.33}}
  \put(69,6.5){$+$}
  \put(75,0){\epsffile{fmgraph.34}}
\end{picture}
\end{center}
\caption{Diagrams leading to a branch singularity at $s=M^2$.  The
solid line denotes a massive Higgs boson, the dashed lines stand for
massless Goldstone bosons.  The cut indicates where the imaginary part
is taken.}
\label{fig:width3}
\end{figure}

In the diagrams in Fig.\ref{fig:width3}, the Higgs ``decays'' by
emitting two massless Goldstone bosons.  Since the ingoing Higgs line
can become off-shell with $\sqrt{s}=M^*>M$, these diagrams give a
contribution with a branch point at $s=M^2$ that merges with the
one-particle pole.  This contribution seems to be of the same order as
$M\Gamma^{(1)}$, potentially invalidating our resummation scheme.
However, at threshold the phase space for the emission of two massless
particles scales like $(s-M^2)^3$, canceling the adjacent Higgs
propagators\footnote{Incidentally, in the sum of the diagrams of
Fig.\ref{fig:width3} the leading term for $s\to M^2$ vanishes in any
representation, contributing another suppression factor $(s-M^2)^2$ to
the $H^*\to Hww$ ``partial width''.}.  Thus, these diagrams are part
of the non-resonant background at two-loop order and are consistently
neglected in a NLO calculation.

\begin{figure}
\begin{center}\leavevmode
\epsffile{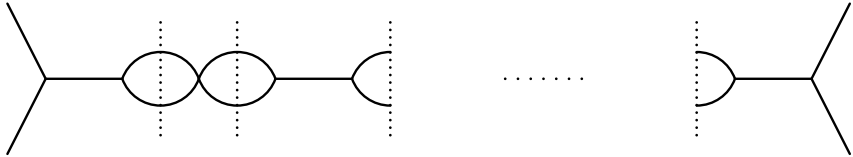}
\end{center}
\caption{Dyson series with one non-resonant insertion}
\label{fig:1L-nonr-chain}
\end{figure}

Another set of (apparently) NLO diagrams consists of those chains
with exactly one non-resonant part (Fig.\ref{fig:1L-nonr-chain}).
Their effect is in fact a two-loop pole shift, which is irrelevant to
the order we are interested in.  The same applies to the imaginary
parts of the diagrams in Fig.\ref{fig:1L-resonant} if they are
accompanied by resonant parts on both sides.  Only in NNLO one has to
be careful that the appropriate definition of the pole mass is
maintained\footnote{This apparent contradiction is resolved by
observing that on the resonance the LO amplitude is purely imaginary
and of order $\lambda^0$.  A pole shift of order $\lambda$ adds a real
part.  In the cross section this amounts to a correction of order
$\lambda^2$.}.

The second relevant set of NLO contributions results from the
imaginary parts of the diagrams in Fig.\ref{fig:1L-resonant} if they
are at one of the ends of the chain.  Including those, the singly
resonant term is given by
\begin{equation}
  A^{(1)}_{\rm res}(s) = 
	-2\lambda\left[1 + \gamma_1
		+ i\theta(s)\frac{\lambda}{16\pi}
		\left(-10 + 4g_1\right)\right]
	\frac{M^2}{s-M^2}.
\end{equation}

\begin{figure}
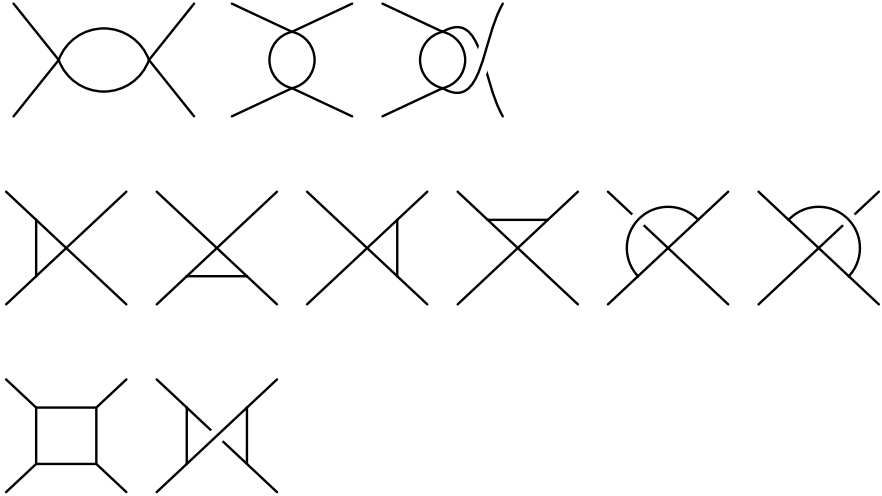

\begin{center}\leavevmode
\unitlength 1mm
\begin{picture}(115,88)
\put(0,0){\epsffile{fmgraph.21}}
\put(20,0){\epsffile{fmgraph.22}}
\put(0,25){\epsffile{fmgraph.15}}
\put(20,25){\epsffile{fmgraph.16}}
\put(40,25){\epsffile{fmgraph.17}}
\put(60,25){\epsffile{fmgraph.18}}
\put(80,25){\epsffile{fmgraph.19}}
\put(100,25){\epsffile{fmgraph.20}}
\put(0,50){\epsffile{fmgraph.23}}
\put(30,50){\epsffile{fmgraph.24}}
\put(50,50){\epsffile{fmgraph.25}}
\end{picture}
\end{center}
\caption{Non-resonant one-loop Feynman diagrams for Goldstone scattering}
\label{fig:1L-nonresonant}
\end{figure}

The non-resonant part consists of the diagrams in
Fig.\ref{fig:1L-nonresonant}
\begin{eqnarray}
  A^{(1)}_{\rm 4pt}(s,t,u) &=& -2\lambda\left\{1 + \frac{\lambda}{16\pi^2}
	\Big[-7\ln|\sigma| - 2\ln|\tau| - 2\ln|\upsilon|
		+ K(\sigma) - 24 \vphantom{\frac12}
   \right.
  + \left(25-9K_1\right)\dos
  \nonumber\\
  && \qquad {}+ 4G(\sigma) + 4G(\tau) + 4G(\upsilon) + 4H(\sigma)
	    + 2F(\sigma,\tau)+2F(\sigma,\upsilon)
  \Big].
  \nonumber\\ &&{}
  +i \frac{\lambda}{16\pi}\Big[
  \left(-7 + 4g(\sigma)\right)\theta(s) 
  \nonumber\\
  && \qquad {}
  + \left(-2 + 4g(\tau) + 2f_1(\sigma,\tau)\right)\theta(t)
  + \left(-2 + 4g(\upsilon) + 2f_1(\sigma,\upsilon)\right)\theta(u)
  \nonumber\\
  && \left.\qquad {}
  + \left(-\beta + 4h(\sigma) 
	+ 2f_2(\sigma,\tau) + 2f_2(\sigma,\upsilon)\right)\theta(s-4M^2)
  \Big]
  \vphantom{\frac{\lambda}{16\pi^2}}\right\}
\end{eqnarray}
and the non-resonant remainder of the diagrams in Fig.\ref{fig:1L-resonant}:
\begin{eqnarray}
  A^{(1)}_{\rm rem}(s) &=& -2\lambda\left\{
	\frac{\lambda}{16\pi^2}
	\left[-2 + \frac{M^4}{(s-M^2)^2}
		\left[-3\ln|\sigma| + 9(K(\sigma)-K_1)	\right]
	- \frac{M^2}{s-M^2}(3-9K_1')  
  \right.\right.\nonumber\\ &&\qquad\left.{} 
	+ \frac{M^2}{s-M^2}
		\left[-10\ln|\sigma| + 6(K(\sigma)-K_1) + 4(G(\sigma)-G_1)
			+ 12(H(\sigma)-H_1) \right]
	\right]
  \nonumber\\ && {}
	+ i\frac{\lambda}{16\pi}
	\left[-9\frac{M^4}{(s-M^2)^2}\beta\theta(s-4M^2)
  \right.\nonumber\\ && \qquad\left.\left.{}
		+ \frac{M^2}{s-M^2}
		\left[4(g(\sigma)-g_1)\theta(s)
			+ 2(-6\beta + 12h(\sigma))\theta(s-4M^2)\right]
	\right]
	\right\}
\end{eqnarray}
The expressions for $A^{(1)}_{\rm 4pt}$ and $A^{(1)}_{\rm rem}$ have
been presented in the linear representation, \emph{i.e.} for $\eta=0$.
However, their sum is independent of $\eta$.

We rewrite the resummed amplitude according to~(\ref{ImpDyson}),
obtaining the final result (still without renormalization-group
improvement):
\begin{eqnarray}\label{NLO}
  A^{(1)}(s,t,u) &=& -2\lambda\left[1 + \gamma_1 
	+ i \theta(s)\frac{\lambda}{16\pi}
		\left(-4+4g_1\right)\right]
	\frac{s}{s-M^2 + iM\Gamma^{(1)}\frac{s}{M^2}\theta(s)}
	\nonumber\\
	&& {}
	+ A^{(1)}_{\rm 4pt}(s,t,u) + A^{(1)}_{\rm rem}(s)
	+ 2\lambda\left[1 + \gamma_1 
		+ i\theta(s)\frac{\lambda}{16\pi}
		\left(-7+4g_1\right)\right].
\end{eqnarray}
Since all tree-level and one-loop contributions are included, their
sum has the correct low-energy behavior.  Any additional $n$-loop
contributions arising from re-expanding the denominator of (\ref{NLO})
are suppressed by $n+1$ powers of $s$.

We have not yet checked the normalization in the low-energy limit.  If
expressed in terms of the renormalized parameters $\lambda$ and $m$,
it reads:
\begin{equation}
  A^{(1)}(s,t,u) \stackrel{s\to 0}{\longrightarrow} 2\lambda\frac{s}{m^2}
	\left[1 - \frac{\lambda}{16\pi^2}(1-\dos)\right]
	+ {\cal O}(s^2\ln s, \cdots).
\end{equation}
Trading $m$ for $M$ according to~(\ref{m-matching}) in this expression,
from the matching condition~(\ref{LE-matching}) we find
\begin{equation}
  \lambda = \frac{M^2}{2v^2}\left[1 + \frac{M^2}{32\pi^2v^2}
		(25-9K_1)(1-\dos)\right].
\end{equation}
Thus the two parameters $\lambda$ and $m$ have been fixed.  In the
on-shell scheme the matching correction vanishes by construction.
Expressed in terms of the physical parameters $G_F=1/(\sqrt{2}\,v^2)$
and $M$, the amplitude $A^{(1)}(s,t,u)$ is scheme-independent up to
the order it has been calculated.

\section{NLO renormalization-group improvement}
\label{sec:twoloop}
The high-energy limit of the amplitude $A^{(1)}$ is not changed by the
resummation in~(\ref{NLO}):
\begin{eqnarray}\label{A1-he}
  A^{(1)}_{\rm he}(s,t,u) &=& -2\lambda\left\{
  1 + \frac{\lambda}{16\pi^2}\left[
  8\ln\frac{|s|}{M^2} + 2\ln\frac{|t|}{M^2} + 2\ln\frac{|u|}{M^2} -26
  + (25-9K_1)\dos
  \right]\right.
  \nonumber\\
	&& \left. \quad{}+ i\frac{\lambda}{16\pi}
	\left[ -8\theta(s) - 2\theta(t) - 2\theta(u)\right]\right\}.
\end{eqnarray}
It should be compared with the one-loop $\MS$ result in the massless theory
\begin{eqnarray}\label{A1-HE}
  A^{(1)}_{\rm HE}(s,t,u) &=& -2\lambda\left\{
  1 + \frac{\lambda}{16\pi^2}\left[
  8\ln\frac{|s|}{\mu^2} + 2\ln\frac{|t|}{\mu^2} 
  + 2\ln\frac{|u|}{\mu^2} -24
  + \dlm\right]\right\}
	\nonumber\\
	&&\left. \quad{}
  + i\frac{\lambda}{16\pi}
  \left[-8\theta(s)-2\theta(t)-2\theta(u)\right]\right\}.
\end{eqnarray}
In the massless theory we have included a matching correction $\dlm$.
Equating the two expressions at a matching point $\mu=\mu_0$ we obtain
its value
\begin{equation}
  \dlm = -2 + 12\ln\frac{\mu_0^2}{M^2} + (25-9K_1)\dos.
\end{equation}
The matching point $\mu_0$ should be taken of the order~$M$.  The
usual choice is $\mu_0=M$, but there are indications~\cite{RW} that 
\begin{equation}\label{improved-mu0}
  \mu_0^2 = \exp(2)\,M^2
\end{equation}
might be a better choice.

The renormalization-group improved result is now obtained from the
high-energy limit as
\begin{equation}\label{A1-RG}
  A^{(1)}_{\rm RG}(s,t,u) = 
  K_\gamma(s) \,A^{(1)}_{\rm HE}(s,t,u) 
  \Big|_{\textstyle \lambda\to \lambda^{(2)}(s)},
\end{equation}
where the two-loop running coupling is given by
\begin{equation}
  \lambda^{(2)}(\mu^2)
  = \frac{\lambda}{1 - \beta_0\lambda\ln\frac{\mu^2}{\mu_0^2}
                     + \frac{\beta_1}{\beta_0}\lambda
                     \ln\left[1 - \beta_0\lambda\ln\frac{\mu^2}{\mu_0^2}
                     \right]}
\end{equation}
with
\begin{equation}
  \beta_0 = \frac{12}{16\pi^2}, \quad
  \beta_1 = -\frac{13}{16\pi^2}\beta_0,
\end{equation}
and the field anomalous dimension gives rise to the factor
\begin{equation}
  K_\gamma(\mu^2) = 1 + \frac{\lambda^{(2)}(\mu) - \lambda}{16\pi^2}
  = 1 + 12\frac{\lambda^2}{(16\pi^2)^2}\ln\frac{\mu^2}{\mu_0^2} + \cdots.
\end{equation}

Thus, the full result~(\ref{amp-full}) at next-to-leading order has
the form
\begin{equation}\label{A1-full}
  A^{(1)}_{\rm full}(s,t,u) = A^{(1)}(s,t,u) 
	+ \left[A^{(1)}_{\rm RG}(s,t,u;\mu_0^2) 
	- A^{(1)}_{\rm HE}(s,t,u;\mu_0^2)\right]
	\theta(s,t,u;\mu_0^2),
\end{equation}
where $A^{(1)}(s,t,u)$ should be taken from~(\ref{NLO}), $A_{\rm RG}$
and $A_{\rm HE}$ are given by~(\ref{A1-HE}) and~(\ref{A1-RG}),
respectively, and $\theta(s,t,u;\mu_0^2)$ is defined in~(\ref{theta}).

\section{The use of representation dependence}
\label{sec:eta}
In the previous sections we have been careful to work only with
representation-independent quantities.  In a practical calculation it
may be useful to take the Feynman rules derived
from~(\ref{sigma-L}, \ref{eta-rep}), and use the fact that the
parameter~$\eta$ drops out of all physical results as a check of the
calculation.  However, it is also instructive to look at
representation dependence from a different viewpoint:

As mentioned in the introduction, in the non-linear representation
($\eta=1$) the power-like low-energy behavior is manifest in each
individual Feynman diagram.  By contrast, at high energies the
complete amplitude rises logarithmically; for individual diagrams this
applies only in the linear representation ($\eta=0$).  In both
representations the respective opposite limit is also reproduced
correctly, but it requires large cancellations between different
Feynman diagrams.  These cancellations would be spoiled by a
na\"\i{}ve introduction of the Higgs boson width.

Therefore, instead of rearranging the perturbation series in the way
described in the preceding sections, one is tempted to interpolate the
two representations by introducing an $s$-dependent $\eta$~parameter.  Of
course, this makes no sense in the Lagrangian, but in a scattering
amplitude $s$ is an external parameter, and since the full amplitude
is independent of $\eta$, this manipulation can only affect
higher-order terms which are not yet included in the perturbative
result.  A natural choice is
\begin{equation}\label{eta(s)}
  \eta(s)= (1+\sqrt{s}/M)^{-1}
\end{equation}
which has the appropriate limits
\begin{equation}
  \eta(0)=1, \qquad
  \eta(\infty)=0.
\end{equation}
Not surprisingly, this exactly reproduces~(\ref{Seymour}) when
inserted into~(\ref{decomp-eta}) with~(\ref{Gamma(eta)}).

It is straightforward to extend this trick to higher orders.  If
working to order $n$, all diagrams with $k$ resonant Higgs propagators
and $n+k$ loops should be evaluated in the general
representation~(\ref{eta-rep}) and resummed.  This requires
computation of the same set of one-particle irreducible diagrams that
are needed in the scheme described before.  If the result is
re-expanded in powers of~$\lambda$, the $\eta$-dependence disappears
up to order~$n$.  However, the full expression is
representation-dependent.  It seems reasonable that by making $\eta$ a
function of $s$, the complete expression fulfils all requirements in
the various energy ranges if the following conditions are satisfied:
\begin{eqnarray}
  1-\eta \propto \phantom{1/}\sqrt{s} &\qquad& \mbox{for $s\to 0$}, \\
  \eta\propto 1/\sqrt{s}              &\qquad& \mbox{for $s\to\infty$}.
\end{eqnarray}
In addition,
the condition 
\begin{equation}
  \eta(M^2)=1/2
\end{equation}
enforces to leading order the kinematical phase space scaling
$\Gamma\propto s$ on the resonance peak.

In this way, a valid formula for the resonance peak can be obtained
from a direct resummation of self-energies.  The disadvantage is that
the parameter~$\eta$ must be kept in the calculation.  However, by
comparing the result for different (reasonable) functions $\eta(s)$
that satisfy the above conditions, the residual representation
dependence from higher orders can be quantified, and the theoretical
uncertainty in the description of the Higgs line shape be estimated.

\section{Discussion}
\label{sec:discussion}
The increase of the running Higgs self-coupling limits the use of
perturbation theory to energies below the Landau pole which arises at
one-loop order in the high-energy limit.  However, within the
perturbative region the procedure described in the present paper is
sufficient for a consistent treatment of the Higgs resonance in the
Goldstone-boson approximation: The physical Higgs mass~$M$ is defined
as the real part of the pole position.  The other free parameter, the
coupling constant~$\lambda$, is fixed by imposing the low-energy
theorem which is a consequence of the custodial $SU_2$ symmetry.  This
matching condition restricts possible parameterizations of the Higgs
resonance and determines the proper inclusion of the Higgs width.  The
resummation of logarithms in the high-energy region can be performed
in a massless theory, and the result can be added to the amplitude
derived in the massive theory in such a way that double-counting is
avoided.  The massless theory has one free parameter which is
determined by a matching condition at a scale $\mu_0\sim M$.  The
final formula~(\ref{amp-full}) describes the Higgs resonance for all
energies, and has been evaluated explicitly to leading (LO) and
next-to-leading order (NLO) in~(\ref{A0-full}) and~(\ref{A1-full}),
respectively.

\begin{figure}
\begin{center}\leavevmode
\epsffile{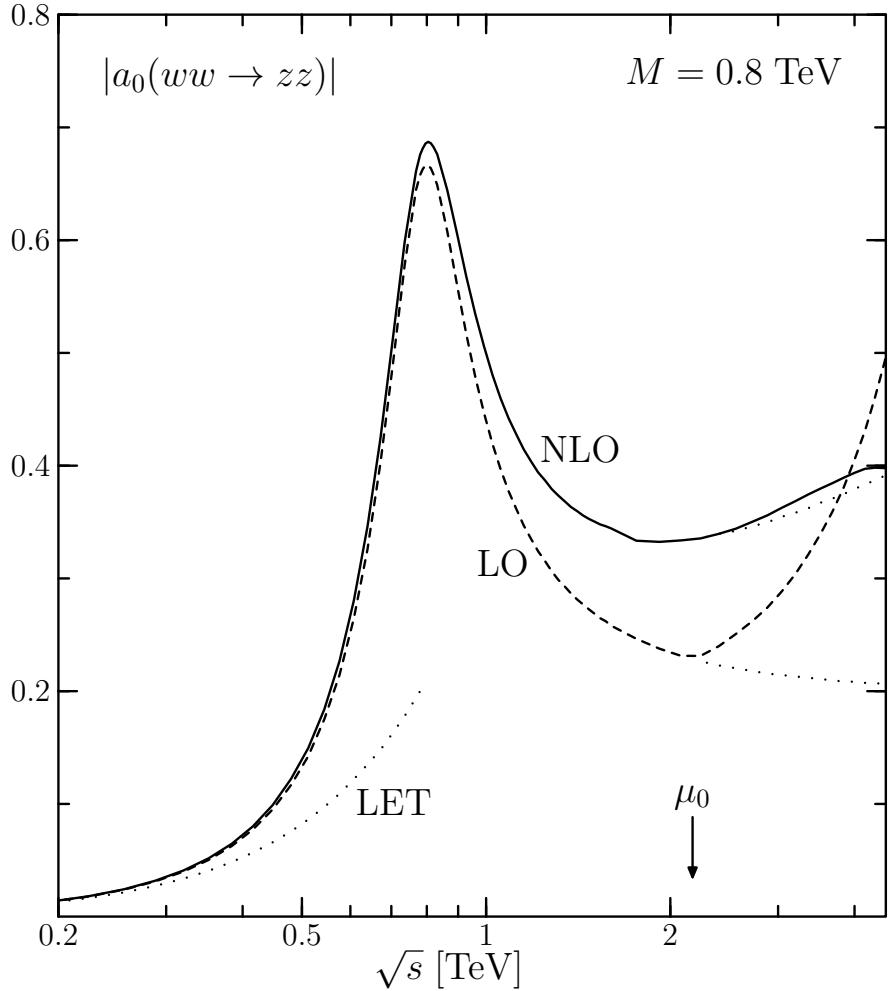}
\end{center}
\caption{Leading-order (LO) and next-to-leading-order (NLO) results
for the Higgs lineshape.  The plot shows the $S$-wave amplitude
$a_0=\protect\frac{1}{16\pi}\int\protect\frac{dt}{s}A(s,t,u)$, using
the formulae (\ref{A0-full}) and (\ref{A1-full}) for $A(s,t,u)$,
respectively.  The low-energy limit~(\ref{LE-matching}) and the
high-energy behavior without renormalization-group improvement are
indicated by dotted lines.}
\label{fig:lineshape}
\end{figure}

The result is shown in Fig.\ref{fig:lineshape} for a Higgs mass
$M=0.8\;{\rm TeV}$.  At low energies, the LO and NLO curves are
virtually indistinguishable: The LO formula~(\ref{A0-full}) already
reproduces the one-loop imaginary part exactly in this limit.  Beyond
the resonance, the LO result rises rapidly towards the Landau pole,
whereas the NLO curve stays at moderate values of the amplitude.  The
transition to the high-energy region at the matching point
$\mu_0=\exp(1)M$ [cf.(\ref{improved-mu0})] is sharp in LO, but smooth
in NLO.

\begin{figure}
\begin{center}\leavevmode
\epsffile{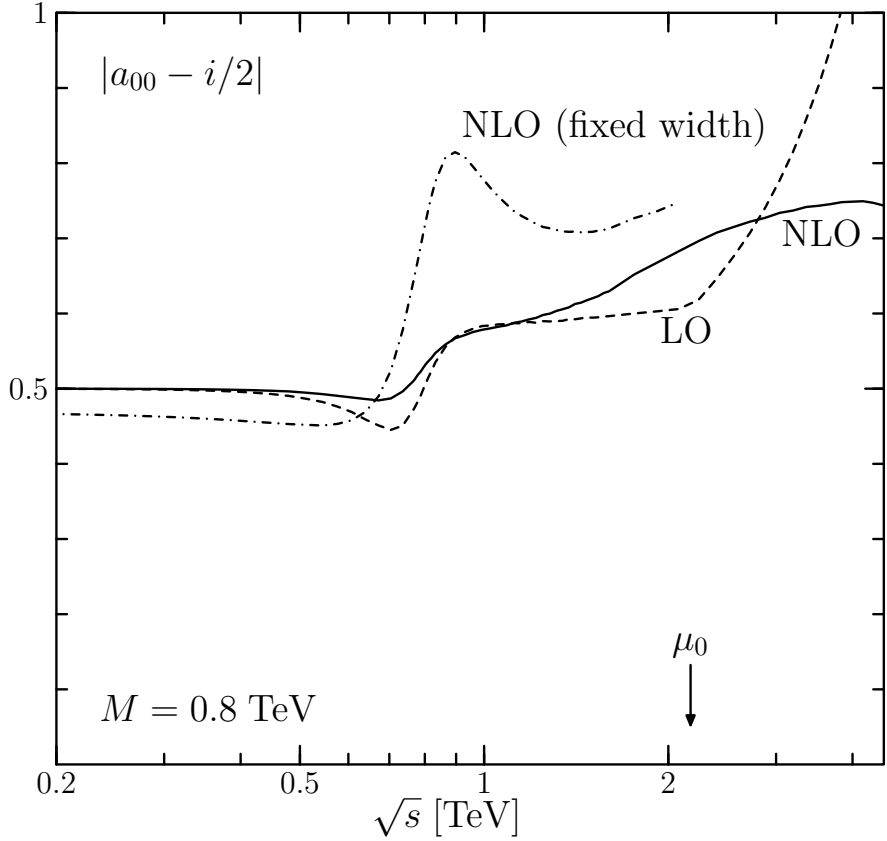}
\end{center}
\caption{Deviation from elastic unitarity, shown for the leading-order
(LO) and next-to-leading-order (NLO) results for the Higgs lineshape,
using the formulae (\ref{A0-full}) and (\ref{A1-full}), respectively.
The partial wave with spin and weak isospin zero is defined as
$a_{00}=\protect\frac{1}{16\pi}\protect\left[\protect\frac{3}{2}\int
\protect\frac{dt}{s}A(s,t,u) +
\int\protect\frac{dt}{s}A(t,s,u)\protect\right]$.  Neglecting
multiparticle thresholds, elastic unitarity requires the quantity
$|a_{00}-i/2|$ to be equal to $1/2$ if all orders are included.  For
comparison, we show the NLO result evaluated according to the
fixed-width formula~(\ref{Dyson}) [dash-dotted line].}
\label{fig:uni}
\end{figure}

To verify that our parameterization is in accordance with unitarity
requirements, in Fig.\ref{fig:uni} we plot the deviation of the
partial-wave eigenamplitude $a_{00}$~\cite{Uni} from the unitarity
circle, the latter given by $|a_{00}-i/2|=1/2$.  Here elastic
unitarity is respected for the formulae~(\ref{A0-full})
and~(\ref{A1-full}) almost perfectly up to $\sqrt{s}=M$, and
approximately up to energies as high as $4\;{\rm TeV}$.  By contrast,
in NLO a fixed-width formula as it directly follows from $S$-matrix
theory (\ref{Dyson}) misses this requirement both at low energies and
in the resonance region, although it is formally equivalent to our
result~(\ref{amp-full}) if all orders are included.  [Note that
higher-order terms will restore unitarity in any scheme which is
consistent in perturbation theory.]

The extension to higher orders is straightforward.  Two-loop
corrections to the Goldstone scattering amplitude and higher-order
renormalization group coefficients have been calculated
in~\cite{Gh,2loop}.

At low energies the transversal degrees of freedom of the gauge bosons
are important, and the gauge couplings and vector boson masses cannot
be neglected.  Although results for physical processes can be obtained
from the Goldstone-boson approximation by means of the effective~$W$
approximation, for numerically reliable predictions the results of
this paper have to be embedded in a full Standard-Model calculation.
In particular, QED bremsstrahlung corrections affect the lineshape and
should be included in conjunction with the process-dependent one-loop
corrections in the electroweak Standard Model~\cite{WW-NLO}.  This
problem will be approached in a future publication.  If a heavy Higgs
resonance has been chosen by Nature for breaking the electroweak
symmetry, it is mandatory to have complete theoretical control over
the lineshape in order to separate effects which could indicate
physics beyond the minimal model.

\section*{Acknowledgments}
The authors are grateful to the theory groups at the Technische
Universit\"at M\"unchen and DESY Hamburg where early parts of this
collaboration were carried out.  Valuable discussions with Th.~Ohl are
also greatly acknowledged.  W.K.\ is supported by German
Bundesministerium f\"ur Bildung und Forschung (BMBF), Contract
Nr.~05~6HD~91~P(0).

\appendix
\section{Integrals}
Here, we define the functions and symbols used in Sec.\ref{sec:NLO}.
In the following abbreviations, $m$~is the mass appearing in the
renormalized propagator; it is equal to the pole mass~$M$ in the
on-shell scheme, or denotes the $\MS$-renormalized mass in the
$\MS$~scheme.  In the one-loop integrals, however, this distinction is
relevant only if the amplitude is evaluated to two-loop order and can
be ignored for the purposes of the present paper.
\begin{equation}
  \sigma = m^2/s, \quad \tau = m^2/t, \quad \upsilon = m^2/u.
\end{equation}
Spence function:
\begin{equation}
  \spence(z) = -\int_0^z \frac{dt}{t}\ln(1-t)
\end{equation}
Two-point integral [$K_1\equiv K(\sigma=1)$]:
\begin{eqnarray}
  K(\sigma) &=& \re\int_0^1 dx\,
		\ln\left(1- \frac{x(1-x)}{\sigma}\right) +2
  \nonumber\\
  &=& \left\{\begin{array}{lc}
		2\sqrt{1-4\sigma}\arcsinh(1/\sqrt{-4\sigma}),
		&\sigma<0,\\
		2\sqrt{1-4\sigma}\arccosh(1/\sqrt{4\sigma}),
		&0\leq \sigma\leq \frac14,\\
		2\sqrt{4\sigma-1}\arcsin(1/\sqrt{4\sigma}),
		&\frac14<\sigma.\end{array}\right.
\end{eqnarray}
Three-point integrals [$G_1\equiv G(\sigma=1)$, $H_1\equiv
H(\sigma=1)$]:
\begin{eqnarray}
  G(\sigma) &=& \sigma\,
		\re\left[\spence\left(\frac{1+\sigma}{\sigma}\right)
			-\frac{\pi^2}{6}\right], \\
  H(\sigma) &=& \sigma\,
		\re\left[\spence\left(\frac{1-\sigma}{x_+-\sigma}\right)
			+\spence\left(\frac{1-\sigma}{x_--\sigma}\right)
			-\spence\left(\frac{\sigma-1}{\sigma}\right)
		\right.\nonumber\\
	    &&  \qquad\left.-\,
			 \spence\left(\frac{-\sigma}{x_+-\sigma}\right)
			-\spence\left(\frac{-\sigma}{x_--\sigma}\right)
			+\frac{\pi^2}{6}\right].
\end{eqnarray}
Four-point integral:
\begin{eqnarray}
  F(\sigma,\tau) &=& \sigma\tau\,
		\re\left\{\frac{1}{\Delta}
		\left[	-\spence\left(\frac{1-z_+}{z_-}\right)
			+\spence\left(-\frac{z_+}{z_-}\right)
		\right.\right.\nonumber\\
		&& \qquad-\,
			 \spence\left(\frac{1-z_+}{x_+-z_+}\right)
			+\spence\left(\frac{-z_+}{x_+-z_+}\right)
			-\spence\left(\frac{1-z_+}{x_--z_+}\right)
		\nonumber\\
		&& \qquad\left.\left.+\,
			 \spence\left(\frac{-z_+}{x_--z_+}\right)
			+\spence\left(\frac{1-z_+}{-z_+}\right)
		-(z_+\leftrightarrow z_-) \right]\right\},
\end{eqnarray}
where
\begin{equation}
  x_\pm = \frac12(1\pm\beta)\quad\mbox{and}\quad
  z_\pm = \frac{1}{2(1+\tau)}(1\pm\Delta),
\end{equation}
and
\begin{equation}
  \beta = \sqrt{1-4\sigma}\quad\mbox{and}\quad
  \Delta = \sqrt{1-4\sigma(1+\tau)}.
\end{equation}
The imaginary parts of the loop integrals are built up by the
functions
\begin{eqnarray}
  g(\sigma) &=& \sigma\ln\frac{1+\sigma}{\sigma},\\
  h(\sigma) &=& 2\sigma\ln\frac{1+\beta}{1-\beta},\\
  f_1(\sigma,\tau) 
    &=& \frac{2\sigma\tau}{\Delta}\ln\frac{\Delta+1}{\Delta-1},\\
  f_2(\sigma,\tau)
    &=& \frac{2\sigma\tau}{\Delta}\ln\frac{\Delta+\beta}{\Delta-\beta}.
\end{eqnarray}
Limiting behavior:
\begin{enumerate}
\item $s\to 0$
\begin{eqnarray}
  K &\to& 2 - \frac16 \frac{s}{m^2} - \frac{1}{60} \frac{s^2}{m^4} + \cdots,
  \\
  G &\to& \left(-\ln\frac{s}{m^2} + 1\right) 
          + \left(\frac12\ln\frac{s}{m^2} - \frac14\right)\frac{s}{m^2}
  \nonumber\\
    && {}+ \left(-\frac13\ln\frac{s}{m^2} + \frac19\right)\frac{s^2}{m^4}
	  + \cdots,
  \\
  H &\to& 1 + \frac{1}{12}\frac{s}{m^2} 
          + \frac{1}{90} \frac{s^2}{m^4} + \cdots,
  \\
  F &\to& \ln\frac{t}{m^2} 
          + \left(-\frac16\frac{s}{m^2}-\frac{t}{m^2}\ln\frac{t}{m^2}\right)
	  + \cdots,
  \\
  g &\to& 1 - \frac12\frac{s}{m^2} + \cdots,
  \\
  f_1 &\to& -1 + t + \cdots.
\end{eqnarray}
\item $s\to M^2$
\begin{eqnarray}
  K &\to& \frac{\pi}{\sqrt3} 
          + \left(1-\frac{2\pi}{3\sqrt3}\right)\frac{s-M^2}{M^2} 
          + \cdots
	\equiv K_1 - K_1'\frac{s-M^2}{M^2} + {\cal O}([s-M^2]^2),
  \\
  G &\to& \frac{\pi^2}{12} + \cdots
	\equiv G_1 + {\cal O}(s-M^2),
  \\
  H &\to& \frac{\pi^2}{9} + \cdots
	\equiv H_1 + {\cal O}(s-M^2),
  \\
  g &\to& \ln 2 + \cdots
	\equiv g_1 + {\cal O}(s-M^2).
\end{eqnarray}

\item $|s|\to\infty$
\begin{eqnarray}
  K&\to& \ln\frac{|s|}{m^2} + {\cal O}(s^{-1}\ln s).
\end{eqnarray}
$G,H,F,g,h,f_1,f_2$ all vanish like $s^{-1}\ln s$ (or faster) in this limit.
\end{enumerate}


\newpage

\newpage

\end{document}